\documentclass[onecolumn]{revtex4}
\usepackage{amsmath,amssymb,graphics,epsfig,subfigure}
\usepackage{color}
\usepackage[colorlinks,linkcolor=red,anchorcolor=red,citecolor=green]{hyperref}
\usepackage{setspace}
\usepackage{booktabs}
\usepackage{float}
\usepackage{appendix}
\usepackage{makecell}
\usepackage{multirow}
\begin{document}

\thispagestyle{empty}

\begin{center}

\title{Generalized Maxwell equal area law and black holes in complex free energy}

\author{Zhen-Ming Xu\footnote{E-mail: zmxu@nwu.edu.cn}, Yu-Shan Wang, Bin Wu, and Wen-Li Yang
        \vspace{6pt}\\}

\affiliation{$^{1}$School of Physics, Northwest University, Xi'an 710127, China\\
$^{2}$Institute of Modern Physics, Northwest University, Xi'an 710127, China\\
$^{3}$Shaanxi Key Laboratory for Theoretical Physics Frontiers, Xi'an 710127, China\\
$^{4}$Peng Huanwu Center for Fundamental Theory, Xi'an 710127, China}

\begin{abstract}
Maxwell equal area law is an important and traditional analytical tool in thermodynamic phase transition, especially in the calculation of gas-liquid phase transition, which reconciles the theoretical calculation with the experimental results. Undoubtedly, its importance is also self-evident for the black hole thermodynamic system. In this study, we construct a generalized Maxwell equal area law, which allows different states of thermodynamic systems to be within the generalized free energy. The black hole thermodynamic characteristics are spontaneously emerged in the free energy landscape. Furthermore, by analytic continuation, we utilize the properties of analytical functions to investigate some universal characteristics of thermodynamic phase transitions in black holes, and preliminarily establish the counterpart of thermodynamic phase transitions in the complex domain.
\end{abstract}

\maketitle
\end{center}

\section{Introduction}
The thermal physics of the black hole benefits from the groundbreaking proposal of the Hawking temperature and Bekenstein-Hawking entropy~\cite{Bekenstein1973,Bardeen1973}, where the temperature of a black hole is proportional to its surface gravity, while the entropy of a black hole is unexpectedly proportional to its event horizon area. Of particular interest is the black hole in the anti-de Sitter (AdS) spacetime, the well-known Hawking-Page phase transition~\cite{Hawking1983} is successfully elaborated as the confinement/deconfinement phase transition in a gauge field~\cite{Witten1998}. Especially in low dimensions, some thermodynamic properties of the black hole can correspond well with that in the condensed matter systems, typically, the correspondence between the Jackiw-Teitelboim gravity and the Sachdev-Ye-Kitaev model~\cite{Maldacena2016,Polchinski2016,Yuan2023}. Thereby, the black hole thermodynamics plays an important role in studying some strongly coupled systems and connecting quantum mechanics, general relativity and statistical physics.

The everyday discussions of black hole thermodynamics are more and more abundant. From the analogy with the van der Waals fluid~\cite{Kastor2009,Dolan2011,Kubiznak2012,Kubiznak2017}, to phenomenological analysis of the microstructure of black hole phase transition~\cite{Wei2015,Miao2018,Wei2019,Xu2020,Ghosh2020}, to the holographic interpretation of black hole thermodynamics~\cite{Zhang2015,Visser2022,Cong2021,Cong2022,Gao2022,Frassino2023}, to the dynamic process of black hole thermodynamics phase transition~\cite{Li2020a,Li2020b,Wei2021,Cai2021,Yang2022,Xu2021a,Xu2021b,Xu2023,Liu2023}, and to the recently proposed topological study of black hole thermodynamics~\cite{Wei2022a,Wei2022b,Wu2023a,Wu2023b,Yerra2022a,Yerra2022b,Zhang2023,Fang2023}, people's research interest has been greatly increased, and the understanding of the microscopic mechanism of the black hole has also been greatly promoted.

Thermodynamics is usually carried out over a span of the real variables. Maxwell equal area law is an important and traditional analytical tool in thermodynamic phase transition, especially in the calculation of gas-liquid phase transition, which reconciles the theoretical calculation with the experimental results. In this study, we construct a generalized Maxwell equal area law, which allows different states of thermodynamic systems to be within the generalized free energy. The standard Maxwell equal area law is just a special case we have constructed so far. The black hole thermodynamic characteristics are spontaneously emerged in the free energy landscape.

On the other hand, the power of the complex function is self-evident. There is a famous saying attributed to the French mathematician Jacques Hadamard ``The shortest path between two truths in the real domain passes through the complex domain.''  The well-known physicist Paul Dirac also emphasized that the theory of functions of a complex variable is an interesting mathematical theory that fulfilled his criteria of beauty, and found this field to be of ``exceptional beauty'' and hence likely to lead to deep physical insight. Hence we have to believe that when we apply complex analysis to the black hole thermodynamics, it will bring unexpected results.

By analytic continuation, we utilize the properties of analytical functions to investigate some universal characteristics of thermodynamic phase transitions in black holes. Often resorting to the complex plane, we preliminarily establish the counterpart of thermodynamic phase transitions in the complex domain, which intuitively and vividly give the picture of the phase transition of the black hole thermodynamic system. We find that the Hawking-Page phase transition is the switch of the combination of two vortices of the vector flow, while the van der Waals-type phase transition is predicted by the local maximum winding number, and the global winding number provides a universal pattern for the thermodynamic phase transition of black holes.

In the following, we first describe how we construct the generalized Maxwell equal area law and briefly explain its characteristics. Then, we extend the analysis of generalized free energy to that in complex domain, discuss the thermodynamic characteristics of black holes on the complex plane, and use the characteristics of analytical functions and {\em Argument Principle} to establish the complex counterpart of thermodynamic phase transitions. At last, we are devoted to providing general conclusions and future prospects. The calculations of some typical thermodynamic systems of black holes are placed in the Appendix. Throughout this paper, we adopt the units $\hbar=c=k_{_{B}}=G=1$.

\section{Generalized Maxwell equal area law}
\begin{figure}[htb]
\begin{center}
\subfigure[~Maxwell equal area law when $\text{area}(A)=\text{area}(B)$]{
		\includegraphics[width=5cm]{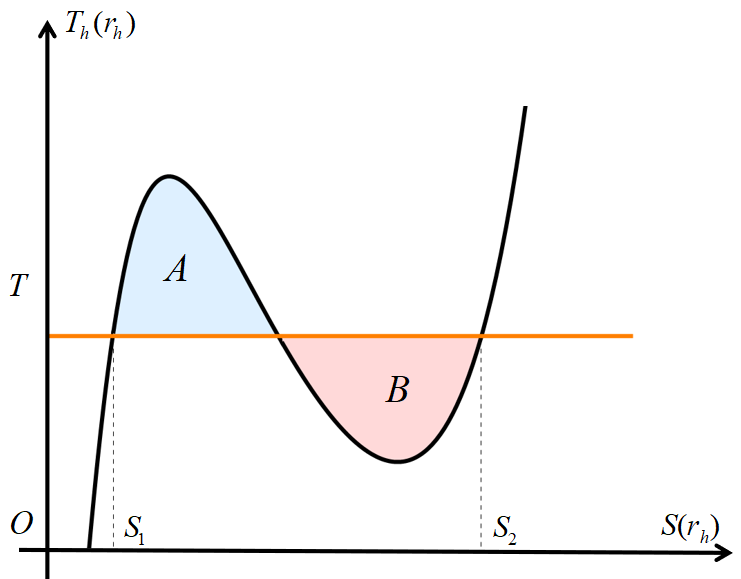}}
\subfigure[~Black hole states in generalized free energy]{
		\includegraphics[width=5cm]{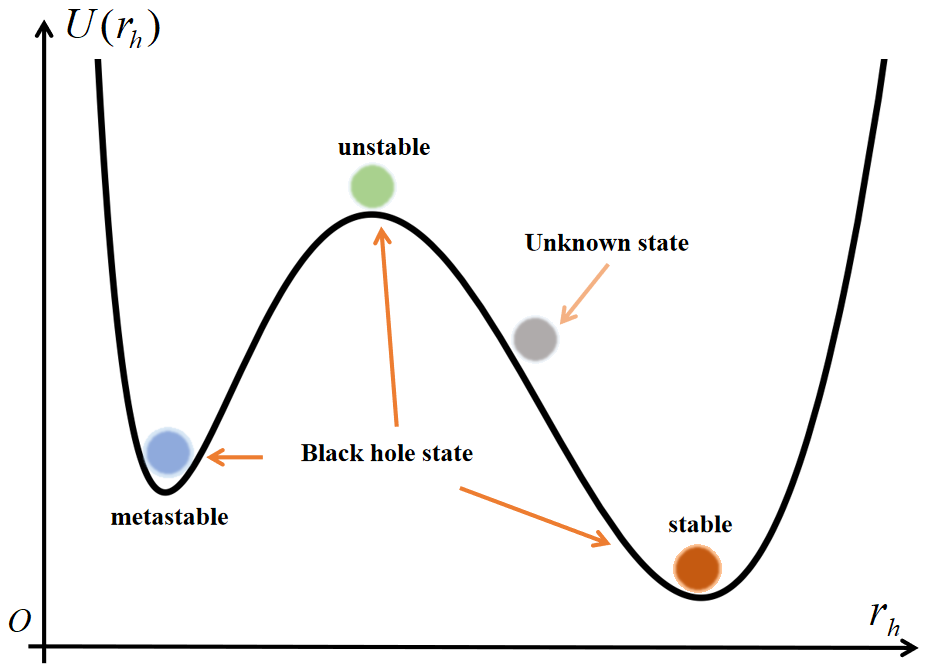}}
\end{center}
\caption{The process of generalized Maxwell equal area law is shown schematically. The diagram (a) is the standard Maxwell equal area law when $\text{area}(A)=\text{area}(B)$. If $\text{area}(A)\neq\text{area}(B)$ and the upper and lower limits of integration are free, then it is the generalized Maxwell equal area law, defined by Eq.~(\ref{potential}), and shown in diagram (b) which presents the behavior of different states of black hole thermodynamic systems in the generalized free energy.}
\label{fig1}
\end{figure}

The Maxwell equal area law is the most important analytical tool in calculations of the thermodynamic phase transition, and its precise solution can provide an analytical representation of the phase diagram. In the diagram (a) in FIG.~\ref{fig1}, we provide a schematic diagram of the construction of the equal area law. In the $\{T_h(r_h),S(r_h)\}$-plane, or $\{P(r_h),V(r_h)\}$-plane, there is an oscillating part when the pressure or temperature is below the critical value, where $T_h(r_h), ~S(r_h), ~P(r_h), ~V(r_h)$ are the Hawking temperature, Bekenstein-Hawking entropy, thermodynamic pressure and volume of the black hole system respectively, and all of these quantities can be written as the functions of the radius of the event horizon $r_h$. Maxwell equal area law tell that this oscillating part can be replaced by an isotherm or isobar in order to describe it in such a way that the areas above and below the isotherm or isobar are equal to each other. Mathematically, the Maxwell equal area law is expressed as in terms of the isotherm $T$ of first-order phase transition in the system,
\begin{eqnarray}\label{earea}
\int_{S_1}^{S_2} (T_h(r_h)-T)\text{d}S(r_h)=0.
\end{eqnarray}

Maxwell equal area law is the most practical method for calculating the first-order phase transition of gas-liquid systems. If we release some of the conditions in this method, so that the various states in the system are in the free energy landscape, some thermodynamic properties of the system will emerge from it, providing useful guidance to understand the first-order phase transition of thermodynamic systems. Note that for the black hole thermodynamic system, its temperature function is $T_h=T_h(P, S)$, where the thermodynamic pressure $P$ is defined by the negative cosmological constant. Based on this, the equation of state of the black hole thermodynamic system can be inversely solved, $P=P(T_h, V)$. Therefore, the construction of the equal area law is equivalent in the $T-S$ plane and the $P-V$ plane. In our current work, we are dealing with problems on the $T-S$ plane. One reason is that the equal area law on the $T-S$ plane can be easily analytically solved. Another reason is that our current proposal is still feasible for black holes without the negative cosmological constant.

Now we construct the generalized Maxwell equal area law. We set the temperature $T$ as a free parameter which is a positive constant and can be assigned in any way, and make the upper and lower limits of integration free. Then the generalized free energy takes the form
\begin{eqnarray}\label{potential}
U(r_h)=\int (T_h(r_h)-T)\text{d}S(r_h).
\end{eqnarray}

In previous study~\cite{Xu2021b}, we have considered a canonical ensemble within the temperature $T$ composed of a large number of states and used the equation~(\ref{potential}) to construct the so-called thermal potential, and used the mathematical structure of the potential function to reflect the thermodynamic properties of the system, providing a new approach for studying the thermodynamic properties of black holes. When the ensemble temperature $T$ is equal to the Hawking temperature $T_h(r_h)$ of the black hole, we think that the ensemble is made up of real black hole states. Otherwise, the ensemble is composed of various unknown states. In addition, the above integration in Eq.~(\ref{potential}) shows that all possible unknown states in the canonical ensemble deviate from the black hole states.

Here in this paper we provide a more reasonable physical explanation for the original construction scheme in ~\cite{Xu2021b}, which essentially reflects the generalization of Maxwell equal area law. If generalized free energy $U(r_h)=0$ and the integration range is from $S_1$ (entropy of liquid-phase) to $S_2$ (entropy of gas-phase), then this construction degenerates into the standard Maxwell equal area law~(\ref{earea}). For nonzero generalized free energy $U(r_h)$, its extreme point exactly satisfies $T=T_h(r_h)$, which means that the system is completely in a black hole state under this condition. Note that $T=T_h(r_h)$ is on-shell, we cannot directly incorporate the condition into the definition~(\ref{potential}) of generalized free energy. In addition, we also note that the characteristics of these extreme points are linked to the thermal stability of thermodynamic systems. If the extremum point is convex, it means that the system is in a thermally unstable state; If the extremum is concave, the system is in a thermally stable state. Of course, when the system is in thermal stability, there are also two situations: metastable and stable states, which are related to the global and local minimum points. These descriptions can be vividly shown in the diagram (b) in FIG.~\ref{fig1}.

Mathematically, we place various states in the black hole thermodynamic system at the extreme points of the generalized free energy. That is to say, various thermal behaviors of the black hole thermodynamic system will be described in the following way,
\begin{eqnarray}\label{zero}
\psi(r_h):=\frac{\text{d}U(r_h)}{\text{d}S(r_h)}=0.
\end{eqnarray}
In this way, we transform the black hole thermodynamic problem into the zeroes of the function $\psi(r_h)$. Our current research has found that the function $\psi(r_h)$ is generally a polynomial. Therefore, in order to analyze the whole picture of the zeroes of polynomials, placing them in the complex domain is the best approach. Hence we here make the functions $U(r_h)$ and $\psi(r_h)$ be the $U(z)$ and $\psi(z)$ with complex continuation via $r_h\rightarrow z:=x+iy$,
\begin{eqnarray}\label{cc}
U(r_h)\quad \xrightarrow{\text{Complex Continuation}} \quad U(z), \qquad \psi(r_h)\quad \xrightarrow{\text{Complex Continuation}} \quad \psi(z).
\end{eqnarray}
Hence physically speaking, the different states of the black hole thermodynamic system correspond to the positive real zeroes of the analytic function $\psi(z)$. However, here we do not require this physical requirement, but rather relax it to the entire complex plane. Next, we apply the above ideas to investigate some universal characteristics of thermodynamic phase transitions in black holes, and preliminarily establish the counterpart of thermodynamic phase transitions in the complex domain. Before that, we need to draw support from some tools in complex analysis.

\section{Complex free energy}
Through complex continuation, we extend the real generalized free energy $U(r_h)$ into a complex one $U(z)$ with the complex argument $z=x+iy$, where $x$ and $y$ are the real and imaginary parts of the complex argument $z$, respectively. In fact, the real part $x$ of a complex variable $z$ corresponds exactly to the horizon radius $r_h$ in the real generalized free energy $U(r_h)$. Hence the complex free energy $U(z)$ can be written as
\begin{eqnarray}
U(z)=u(x,y)+iv(x,y),
\end{eqnarray}
where $u(x,y)$ and $v(x,y)$ are the real and imaginary parts of the complex free energy $U(z)$. Mathematically, the complex free energy $U(z)$ can represent an energy flow vector field. The overall behavior of this vector field can reflect some thermodynamic characteristics of the system.

\begin{figure}[htb]
\begin{center}
\includegraphics[width=65 mm]{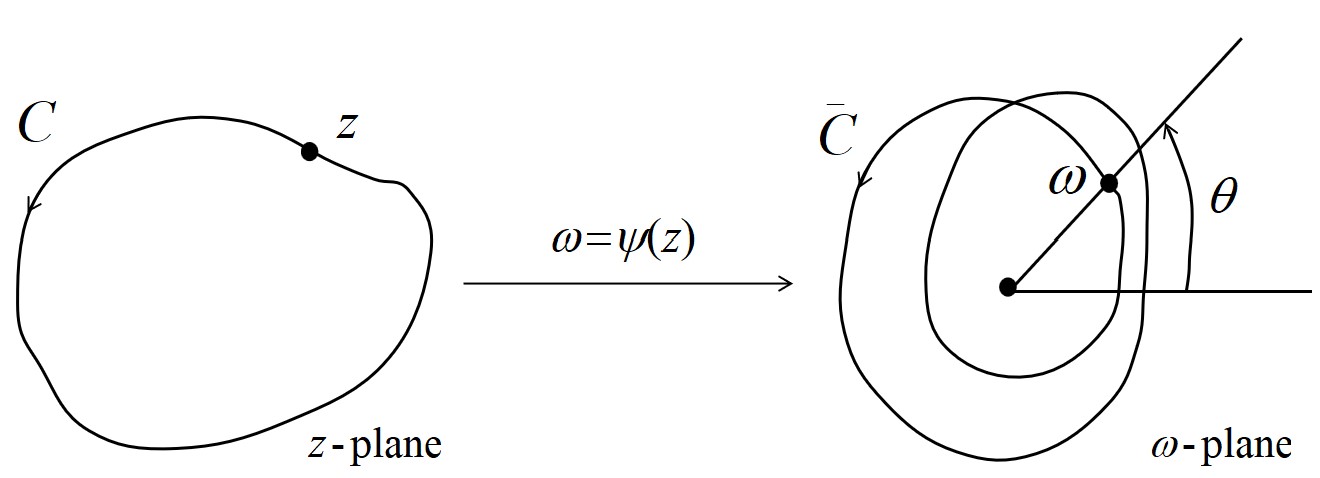}
\end{center}
\caption{Graphic representation of the winding number. Let $\omega$ be the image of the point $z$ under the mapping $\omega=\psi(z)$, and $\theta=\text{arg}\psi(z)$ be the angle that the ray from the origin to $\omega$ makes with respect
to the horizontal.}
\label{fig2}
\end{figure}

In complex analysis, the number of zeroes and poles of a meromorphic function $\psi(z)$ are related to the contour integral. There is an important {\em Argument Principle}. Let $\psi(z)$ be a meromorphic function defined inside and on a simple closed contour $C$, with no zeroes or poles on $C$. Then we have
\begin{eqnarray}\label{argument}
N(\psi,C)-P(\psi,C)=\frac{1}{2\pi i}\oint_{C}\frac{\psi^{'}(z)}{\psi(z)}\text{d}z=\frac{\Delta_{_C}\text{arg}\psi(z)}{2\pi},
\end{eqnarray}
where $N(\psi,C)$ and $P(\psi,C)$ are the number of zeroes and poles, respectively, of $\psi(z)$ inside $C$, and $\psi^{'}(z)$ represents the first derivative of $\psi(z)$ with respect
to the complex variable $z$; where a multiple zero or pole is counted according to its multiplicity, and where $\text{arg}\psi(z)$ is the argument of $\psi(z)$; that is, $\psi(z)= |\psi(z)| \exp(i~\text{arg}\psi(z))$ and $\Delta_{_C}\text{arg}\psi(z)$ denotes the change in the argument of $\psi(z)$ over $C$. Equation~(\ref{argument}) corresponds to the number of times the point $w$ winds around the origin on the image curve $\bar{C}$ when $z$ moves around $C$ under the transformation $w=\psi(z)$, shown in FIG.~\ref{fig2}.

Hence the winding number of $\bar{C}$ about the origin is
\begin{eqnarray}
W:=\frac{1}{2\pi i}\oint_{\bar{C}}\frac{\text{d}\omega}{\omega}=\frac{1}{2\pi i}\oint_{C}\frac{\psi^{'}(z)}{\psi(z)}\text{d}z.
\end{eqnarray}
Specially, there are no poles of $\psi(z)$ inside $C$, i.e., $P(\psi,C)=0$, we have $W=N(\psi,C)$. When the complex variable $z$ changes over the contour $C$, the argument function $\theta=\text{arg}\psi(z)$ may be a multi-valued function. In order to turn a multi-valued function into a single valued one, a common method is to construct the Riemann surface of a multi-valued function. On each foliation of Riemann surface, a complex function is single-valued. According to the meaning of the winding number, when the winding number is one, it corresponds to a single foliation Riemann surface, and when the winding number is two, it corresponds to the Riemann surface with two foliations. In short, the winding number is directly related to the foliations of the Riemann surface of the complex variable function.

Next, we focus on exploring the phase transition characteristics of black hole thermodynamics using complex analysis methods and establish some universal properties of phase transitions.

\subsection{Hawking-Page phase transition}
The most typical Hawking-Page phase transition occurs in the Schwarzschild-AdS black hole. From the perspective of Gibbs free energy, the state with the global minimum of Gibbs free energy is thermodynamically preferred respectively. Through the Hawking radiation process, a stable large black hole can exchange energy and establish the equilibrium with the thermal AdS background, which is considered as the Hawking-Page phase transition. Now we turn to the current perspective of the complex free energy.
\begin{figure}[htb]
	\centering
	\subfigure[~$0<T<T_{\text{HP}}$]{
		\includegraphics[width=5cm]{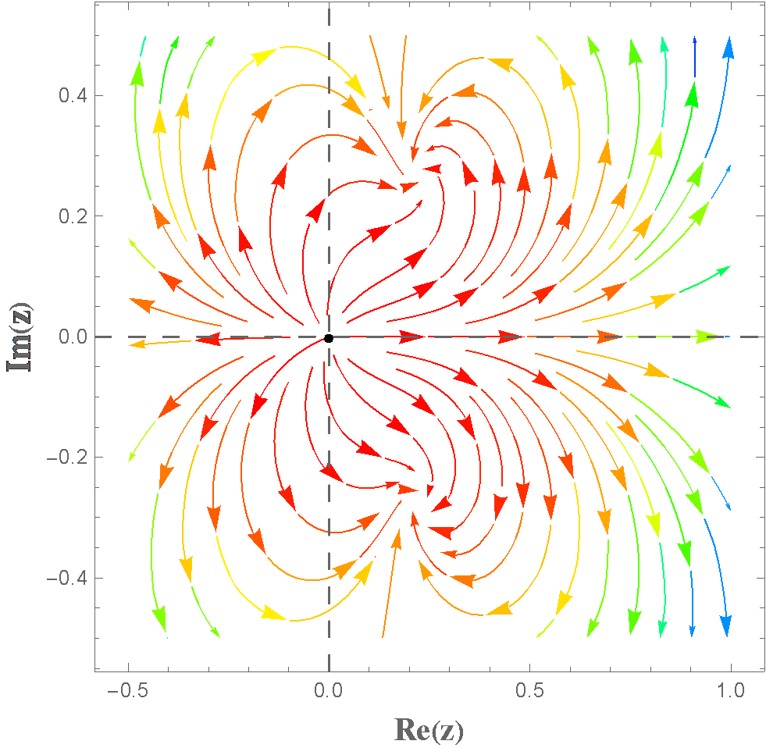}
		\includegraphics[width=5cm]{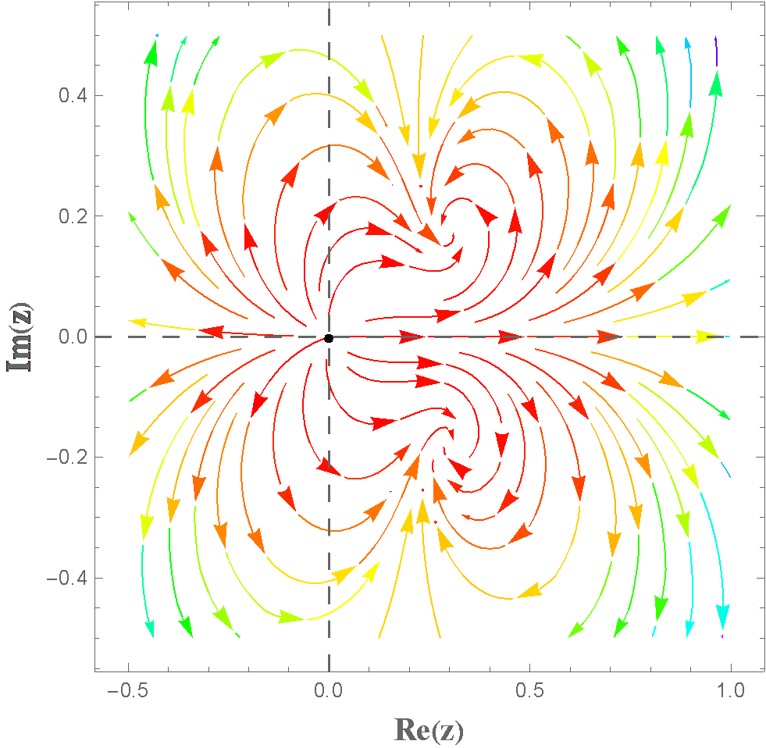}
		\includegraphics[width=5cm]{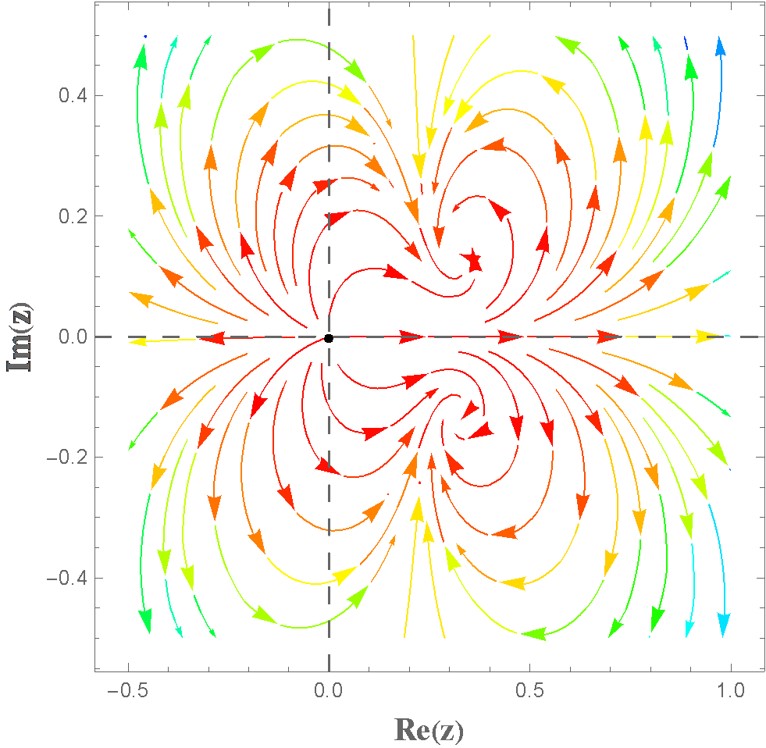}}
	\subfigure[~$T=T_{\text{HP}}$]{
		\includegraphics[width=5cm]{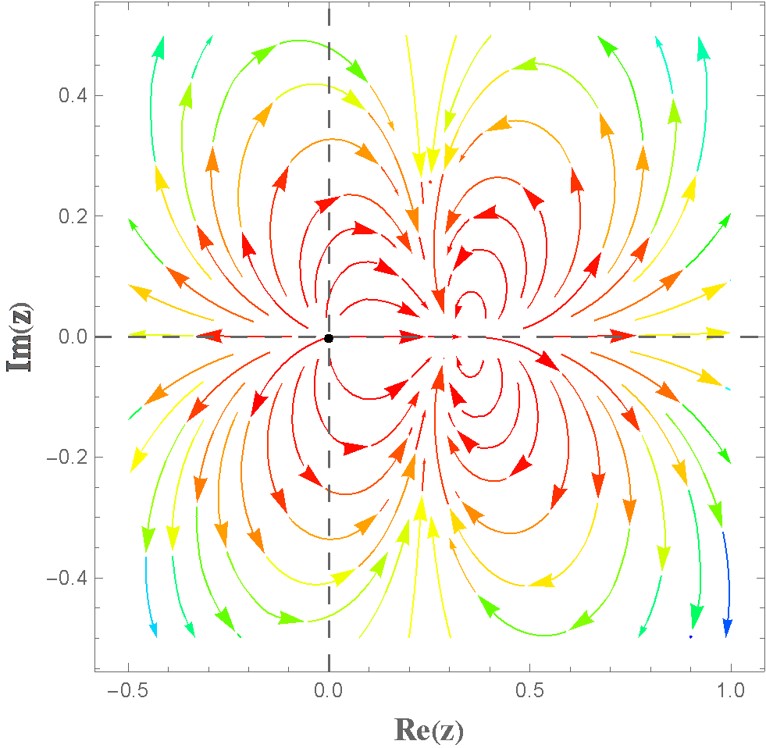}}
	\subfigure[~$T>T_{\text{HP}}$]{	
		\includegraphics[width=5cm]{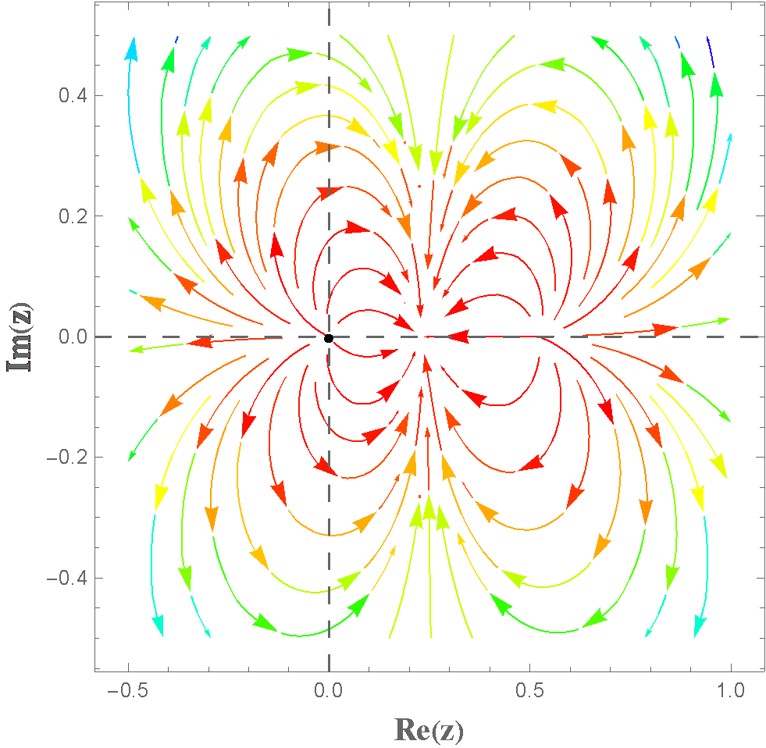}
		\includegraphics[width=5cm]{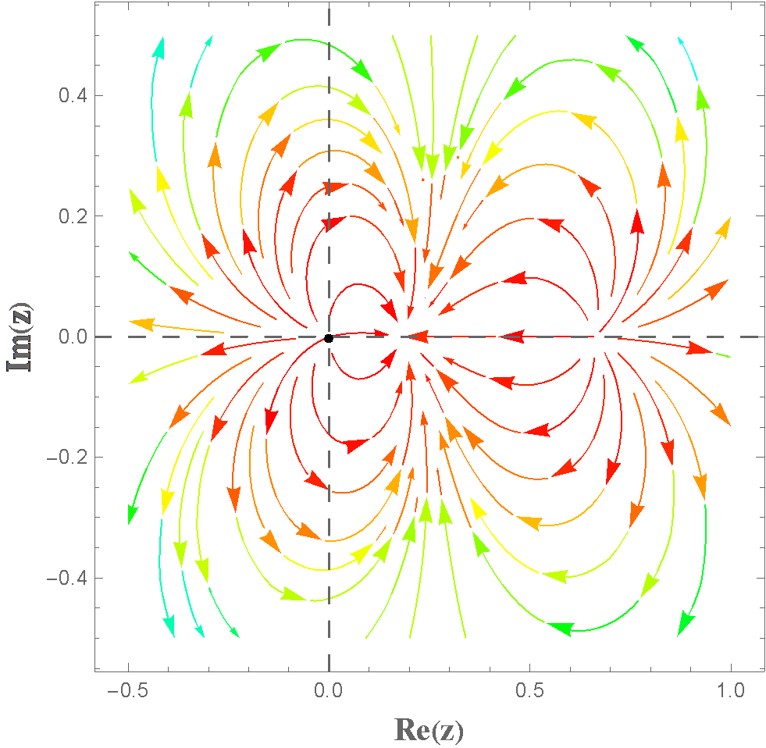}}
	\caption{The energy flow vector plots as the temperature gradually increases for the Schwarzschild-AdS black hole.}
	\label{fig3}
\end{figure}

For the Schwarzschild-AdS black hole, the complex free energy can be obtained through the following methods. The Hawking temperature $T_h(r_h)$ and the Bekenstein-Hawking entropy $S(r_h)$ of the Schwarzschild-AdS black hole are
\begin{eqnarray}
T_h(r_h)=\frac{1}{4\pi r_h}+\frac{3r_h}{4\pi l^2}, \qquad S(r_h)=\pi r_h^2,
\end{eqnarray}
where $l$ is the AdS radius. By substituting the above two equations into Eq.~(\ref{potential}) and completing the integration, and then extending to the complex domain (Eq.~(\ref{cc})), we can have
\begin{eqnarray}\label{sadsu}
U(z)=\frac{1}{2}z-\pi T z^2+\frac{z^3}{2l^2}.
\end{eqnarray}

When $U(z)=0$ and $U'(z)=0$, we can obtain $T=T_{\text{HP}}=\sqrt{1/(\pi l)}$, which is just the Hawking-Page phase transition temperature of the Schwarzschild-AdS black hole. Hence the Hawking-Page phase transition corresponds to the second-order zero of the analytic function $U(z)$.

Next we present an image of the second-order zero of the complex free energy $U(z)$ to facilitate a clear understanding of the Hawking-Page phase transition in the complex plane. The complex free energy $U(z)$ can represent an energy flow vector field. Hence we can write the real part $u(x,y)$ and imaginary part $v(x,y)$ of the complex free energy,
\begin{eqnarray}
u(x,y)&=&\frac{1}{2}x-\pi T(x^2-y^2)+\frac{1}{2l^2}(x^3-3xy^2),\nonumber \\
v(x,y)&=& y\left[\frac12-2\pi T x+\frac{1}{2l^2}(3x^2-y^2)\right].
\end{eqnarray}
Clearly when setting $y=0$, we can have $v(x,y)=0$ and $u(x)=\frac{1}{2}x-\pi T x^2+\frac{x^3}{2l^2}$, which is real generalized free energy. Since the complex function represents a vector field, we have shown the overall behavior of the energy flow vector of the black hole as the temperature gradually increases in FIG.~\ref{fig3}. It is clear that at this point, the main factor is the Hawking-Page phase transition temperature $T_{\text{HP}}$. When the temperature of the black hole is lower than $T_{\text{HP}}$, the vector flow forms vortex structures in the upper half-plane and the lower half-plane of the complex plane respectively; when the temperature of a black hole is equal to the $T_{\text{HP}}$, two vortices begin to merge into one. Hence from the perspective of the complex analysis, the Hawking-Page phase transition is the switch of the combination of two vortices of the energy flow vector. In other words, the Hawking-Page temperature plays a crucial role in the splitting process of vortex structures of the energy flow vector. It is a very interesting result, and its deeper physical correspondence requires further research in the future.

\subsection{van der Waals-type phase transition}
Then we focus on the van der Waals-type phase transition of black holes, such as the heat capacity divergence and the intersection of the Gibbs free energy swallow tail structure. According to Ehrenfest's classification of phase transitions, these are basically first-order or second-order phase transitions. From our current complex analysis point of view, these phase transitions are related to the zeroes of the analytic function $\psi(z)$ defined by Eq.~(\ref{zero}). Based on the {\em Argument principle}, we can establish a correspondence that the local maximum winding number predicts the van der Waals-type phase transition, and the global winding number provides a universal pattern for the thermodynamic phase transition of black holes, and also obtain the counterpart of thermodynamic phase transitions in the complex domain, summarized in TABLE~\ref{tab}. The specific details of the calculation can refer to the Appendix~\ref{app}.
\begin{table}[htb]
\centering
\begin{tabular}{c|c|c|c|c}
\hline
  \multirow{2}{*}{Black holes} & \multirow{2}{*}{Phase transition}& \multicolumn{2}{c|}{Winding number} & \multirow{2}{*}{\makecell[c]{Complex counterpart\\(Riemann surface)}}\\
\cline{3-4}
 & &local-max & ~global~  \\
\hline
\rule{0pt}{22pt} Schwarzschild & No & 1 & 0&one foliation {\includegraphics[width=8 mm]{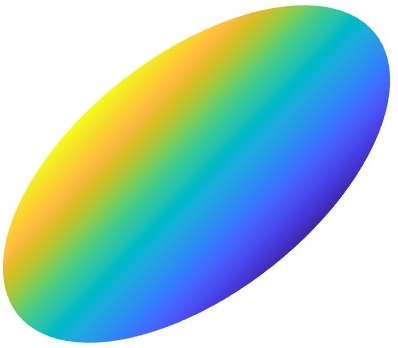}}  \\
\rule{0pt}{22pt} Reissner-Nordstr\"{o}m& $2^{\text{nd}}$ & 2 & 0& two foliations {\includegraphics[width=8 mm]{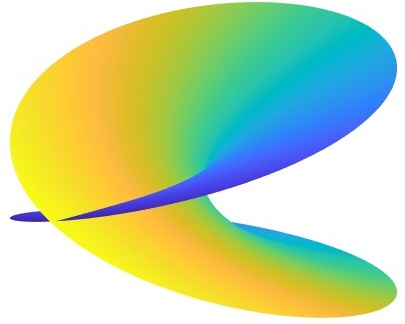}}  \\
\rule{0pt}{22pt} Schwarzschild-AdS &$2^{\text{nd}}$ & 2 & 1&two foliations {\includegraphics[width=8 mm]{two}}  \\
\rule{0pt}{22pt} Charged AdS  &$1^{\text{st}}$ and $2^{\text{nd}}$ & 3 &1 &three foliations {\includegraphics[width=8 mm]{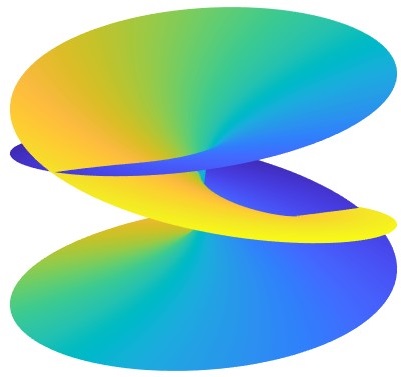}}\\
\rule{0pt}{22pt} 6D charged Gauss-Bonnet&$1^{\text{st}}$ and $2^{\text{nd}}$ & 5 &1 &five foliations {\includegraphics[width=8 mm]{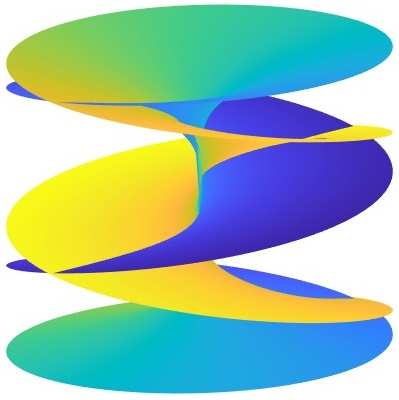}} \\
\hline
\end{tabular}
\caption{The local maximum winding number, the global winding number and the complex counterpart of the thermodynamic phase transition for several typical black holes.}
\label{tab}
\end{table}

We know that the phase transition is related to the zeroes of the analytic function $\psi(z)$. For different black hole models, their corresponding analytical functions $\psi(z)$ are also different. Here we only list a few typical thermodynamic systems of black holes, and thus summarize an empirical criterion. We obtain the local maximum winding numbers of various black holes, where we (i) exclude by hand the poles at $z=0$ by considering the complex plane without the origin, and (ii) consider only zeros that are real and positive. This is because the real and positive $z$ corresponds to physical values for the radius $r_h$. On the other hand, without these artificial restrictions, we can obtain the global winding numbers.

For the local maximum winding number, by decomposing it, we can endow the phase transition with concrete and vivid complex counterpart. In a nutshell, there are three basic elements: (i) when the winding number is one, there is no phase transition, and the corresponding complex counterpart is the Riemann surface with one foliation; (ii) the winding number is two, which corresponds to the second-order phase transition and has a Riemann surface with two foliations; (iii) when the winding number is three, it means that the first-order phase transition will occur, accompanied by the second-order phase transition, which has a Riemann surface with three foliations. Moreover if a black hole system is similar to the water, in which the solid, liquid, and gas phases can coexist, it possesses a triple point. At this time, we observe that such a black hole thermodynamic system with a triple point has a complex counterpart of the Riemann surface with five foliations. The complex counterpart behind the phase transition will pierce our new understanding of the microscopic mechanism of the black hole.

For the global winding number given by the zeros-minus-poles count $N-P$, it reflects a universal characteristic of phase transition in black hole thermodynamic systems. For the Reissner-Nordstr\"{o}m black hole, we are considering the divergence of the heat capacity at fixed charge, which corresponds to a second-order phase transition. For the Schwarzschild-AdS black hole, we also treating the divergence of the heat capacity at constant pressure as the second-order phase transition. The phase transition analysis of other black holes is basically obtained based on the Gibbs free energy.

For the complex counterpart of the thermodynamic phase transition, i.e., Riemann surface, is only based on the local maximum winding number, in order to visually understand phase transitions from a mathematical perspective. Figuratively speaking, a thermodynamic system must have an intermediate transition state if it undergoes a first-order phase transition, so its mathematical structure must be at least three foliations structure. For the second-order phase transition, it is a jump, similar to the effect of dividing one into two, so it must have a double foliations structure. However, the deeper physical implications behind these complex counterpart require further research in the future.

\section{Discussion and Summary}\label{sec4}
In this study, we generalize the Maxwell equal area law, and construct the generalized free energy, which allows different states of thermodynamic systems to be within it. The black hole thermodynamic characteristics are spontaneously emerged in the free energy landscape. Then by analytic continuation, we utilize the properties of analytical functions to investigate some universal characteristics of thermodynamic phase transitions in black holes. For the AdS black hole, there are two type phase transitions, i.e., the Hawking-Page phase transition and van der Waals-type phase transition.

For the Hawking-Page phase transition, it corresponds to the second-order zero of the complex free energy $U(z)$. The Hawking-Page phase transition temperature $T_{\text{HP}}$ is regarded as the switch of the combination of two vortices of the vector flow. For the van der Waals-type phase transition, we relate it to the zeroes of the analytic function $\psi(z)$ defined by Eq.~(\ref{zero}). By grace of an important theorem in analytic functions, i.e., {\em Argument Principle}, we can establish a correspondence that the local maximum winding number predicts the van der Waals-type phase transition, and the global winding number provides a universal pattern for the thermodynamic phase transition of black holes, and also obtain the counterpart of thermodynamic phase transitions in the complex domain. This correspondence is just based on the analysis of several typical black hole thermodynamics systems. The general analysis of this correspondence is already a challenging and open question.

Based on the current results, interestingly, if the local maximum winding number is $W=4$, it appears in the third order Lovelock black holes and will be discussed in detail in the study~\cite{Wang2023}. At this point, we have two decomposition methods: (i)$4=2+2$, it indicates that only the second-order phase transition occurs in the system; (ii)$4=1+3$, it means that the system will have both second-order and first-order phase transitions. These results are consistent with our understanding of the thermodynamic characteristics of the third order Lovelock black holes.

For black holes, all its thermodynamic quantities can be written as a function of the horizon radius $r_h$, therefore it is regarded as a fundamental variable. If we want to returns to the realm of general thermodynamics, the entropy $S$ is the fundamental variable. We know $r_h$ is just the square root of the entropy $S$ and we can use $r_h \propto \sqrt{S}$ instead of $S$ itself that does lead to an interesting property, namely that the functions $U(z)$ and $\psi(z)$ become rational, which enables our current proposed approach of counting zeros and poles. This may indeed be reflected the interesting insight about general thermodynamic systems.

We hope that the current work can provide a new concept for analyzing black hole thermodynamics and even black hole physics. We can further explore the microstructure of the black hole system through the relevant methods in complex analysis, so that we can more clearly glimpse the overall structure of the black hole thermodynamic system. In addition, for the black hole thermodynamic system, it is generally known that there will be a first-order phase transition or a second-order phase transition, or a mixture of the two, and individual systems will show a zeroth-order phase transition. We currently focus mainly on first-order and second-order phase transitions, while the complex counterpart of other order phase transitions require further investigation.

\section*{Acknowledgments}
This research is supported by National Natural Science Foundation of China (Grant No. 12105222, No. 12275216, and No. 12247103). The author would like to thank the anonymous referee for the helpful comments that improve this work greatly.

\appendix

\section{Several typical black holes in complex free energy}\label{app}
\subsection{Schwarzschild black hole}
For a simplest four-dimensional static spherically symmetric black hole, that is Schwarzschild black hole, its Hawking temperature and the Bekenstein-Hawking entropy are $T_h(r_h)=\frac{1}{4\pi r_h}$ and $S(r_h)=\pi r_h^2$. Substitute them into Eqs.~(\ref{potential}) and~(\ref{zero}), and then perform complex continuation (Eq.~(\ref{cc})) to obtain the analytical function
\begin{eqnarray}
\psi(z)=\frac{1}{4\pi z}(1-4\pi T z).
\end{eqnarray}
In the entire complex plane excluding the origin ($\textbf{C}\backslash \{0\}$), this analytical function has only one zero point with first-order. Hence the local maximum winding number of $\bar{C}$ about the origin is $W_{\text{local-max}}=1$, resulting that its complex counterpart is the Riemann surface with single foliation. This results mean that there will be no phase transition in the black hole thermodynamic system. This is also consistent with our understanding of the Schwarzschild black hole thermodynamic system. For the global winding number, it is calculated by the zeros-minus-poles count $N-P$ in the entire complex plane. Hence we have $W_{\text{global}}=1-1=0$.
\subsection{Reissner-Nordstr\"{o}m black hole}
For the charged four-dimensional static spherically symmetric black hole, its Hawking temperature and the Bekenstein-Hawking entropy are $T_h(r_h)=\frac{1}{4\pi r_h}-\frac{Q^2}{4\pi r_h^3}$ and $S(r_h)=\pi r_h^2$. Substitute them into Eqs.~(\ref{potential}) and~(\ref{zero}), and then perform complex continuation (Eq.~(\ref{cc})) to obtain the corresponding analytical function
\begin{eqnarray}
\psi(z)=\frac{1}{4\pi z^3}(z^2-Q^2-4\pi T z^3),
\end{eqnarray}
where $Q$ is the total charge of the black hole. Easily, we can see that the analytical function has two zeroes (real and positive) at most in the complex plane $\textbf{C}\backslash \{0\}$, within the appropriate parameter space. When $T=1/(6\sqrt{3}\pi Q)$, there is only one zero with second-order. When $T<1/(6\sqrt{3}\pi Q)$, there are two zeroes with first-order. Hence the local maximum winding number is $W_{\text{local-max}}=2$, and its complex counterpart is considered as the Riemann surface with two foliations. We already know that the heat capacity of a charged black hole will exhibit divergent behavior, which means that the system will undergo a second-order phase transition. In this way, when the winding number is two, it implies the system undergoing a second-order phase transition, whose the complex counterpart is the Riemann surface with two foliations. While for the global winding number, we have $W_{\text{global}}=3-3=0$.
\subsection{Schwarzschild-AdS black hole}
For the black hole in the AdS background, the simplest static spherically symmetric one is the Schwarzschild-AdS black hole. Based on Eqs.~(\ref{zero}) and~(\ref{sadsu}), its analytical function can be read as
\begin{eqnarray}
\psi(z)=\frac{1}{4\pi z}\left(1+\frac{3z^2}{l^2}-4\pi T z\right).
\end{eqnarray}
Similarly, the analytical function has two zeroes (real and positive) at most in the complex plane $\textbf{C}\backslash \{0\}$, within the appropriate parameter space. When $T=T_{\text{min}}=\sqrt{3}/(2\pi l)$, there is only one zero with second-order. When $T>T_{\text{min}}$, there are two zeroes with first-order. Hence the local maximum winding number is $W_{\text{local-max}}=2$, and its complex counterpart is the Riemann surface with two foliations. This result implies that there is a second-order phase transition in the system. Similar to the results of the Reissner-Nordstr\"{o}m black hole, the heat capacity of the Schwarzschild-AdS black hole also exhibits divergence, indicating that it will undergo a second-order phase transition. Therefore, it once again confirms our understanding of the complex counterpart on second-order phase transition. While for the global winding number, we have $W_{\text{global}}=2-1=1$.
\subsection{Charged AdS black hole}
Then, we turn to the typical example of black hole thermodynamics phase transition, i.e., the charged AdS black hole. Based on the above discussion, the analytical function of the black hole system can be directly obtained as
\begin{eqnarray}
\psi(z)=\frac{1}{4\pi z^3}\left(z^2+\frac{3z^4}{l^2}-Q^2-4\pi T z^3\right),
\end{eqnarray}
where it has three zeroes (real and positive) at most in the complex plane $\textbf{C}\backslash \{0\}$, within the appropriate parameter space. Hence the local maximum winding number is $W_{\text{local-max}}=3$, and its complex counterpart is the Riemann surface with three foliations. For the charged AdS black hole, its thermodynamic behavior is similar to the one of the van der Waals fluids, with both second-order and first-order phase transitions. Thus it can be seen that a thermodynamic system will undergo a first-order phase transition, and the corresponding winding number is at least three, and if a second-order phase transition is to occur, the winding number should be at least two. This observation can be confirmed by the above some examples. While for the global winding number, we have $W_{\text{global}}=4-3=1$.

\subsection{Charged Gauss-Bonnet black hole in six-dimensions}
For the charged Gauss-Bonnet black hole in six-dimensions~\cite{Wei2021,Frassino}, we have known that it is similar to the water, in which the solid, liquid, and gas phases can coexist. The black hole hold a triple point being a coexistence phase of stable small, intermediate, and large black hole states. With a little calculation, we can obtain its analytical function as
\begin{eqnarray}
\psi(z)=\frac{\frac{10z^8}{l^2}+6z^6+2\alpha z^4-Q^2-8\pi Tz^5(z^2+2\alpha)}{8\pi z^5(z^2+2\alpha)},
\end{eqnarray}
where $\alpha$ is effective Gauss-Bonnet coupling constant with positive value and it is regarded as the inverse string tension. Through tedious analysis, it can be concluded that this analytical function can have up to five positive real zeros throughout the complex plane $\textbf{C}\backslash \{0\}$ within the appropriate parameter space. Therefore the local maximum winding number is $W_{\text{local-max}}=5$, and its complex counterpart is the Riemann surface with five foliations. We can decompose the winding number as $5=3+2$, implying that there is one first-order and two second-order phase transitions in the system. This is consistent with our understanding of the phase transition behaviors of the black hole thermodynamic system. It is the black hole thermodynamic phase transition at a triple point. While for the global winding number, we have $W_{\text{global}}=8-7=1$.

\end{document}